\documentclass[twocolumn,showpacs,preprintnumbers,amsmath,amssymb]{revtex4}
\usepackage{dcolumn}
\usepackage{bm}
\begin{document}
\title{Probing vortices in $^4$He nanodroplets}
\author{Francesco Ancilotto}
\affiliation{
INFM (Udr Padova and DEMOCRITOS National Simulation Center,
Trieste, Italy) and
Dipartimento di Fisica ``G. Galilei'',
Universit\`a di Padova, via Marzolo 8, I-35131 Padova, Italy}
\author{Manuel Barranco and Mart\'{\i} Pi}
\affiliation{
Departament E.C.M., Facultat de F\'{\i}sica,
Universitat de Barcelona, E-08028, Spain}

\begin{abstract}

We present static and dynamical properties of linear
vortices in $^4$He droplets obtained from Density
Functional calculations.
By comparing the adsorption properties of different
atomic impurities embedded in pure droplets and in
droplets where a quantized vortex has been created,
we suggest that Ca atoms should be the dopant of choice
to detect vortices by means of spectroscopic experiments.

\end{abstract}


\maketitle

The unique environment realized in liquid $^4$He
clusters has opened up in recent years new
opportunities for atomic/molecular spectroscopy to probe
superfluid phenomena on the atomic scale \cite{toennies,kwon}.
Helium droplets represent ideal
nano-scale cryostats for a variety of fundamental
experiments on liquid $^4$He, including the study of
quantized vortices \cite{close}. Vortices, while
energetically unfavorable \cite{bauer}, can potentially be
stabilized by atomic or molecular impurities \cite{barranco1}.

During the free jet expansion experiments described in
Refs. \cite{toennies,close},
it is plausible that quantized vortices may be created
in some metastable state, long-lived enough to be detected.
However, the question of whether $^4$He droplets
can sustain vortices is still not resolved, and
all the high resolution spectra of embedded molecules
can apparently be
explained without invoking their presence. Yet,
it is expected that in the near future they could be
created by some extension of the present experimental techniques.
This calls for identifying signatures
that might reveal vortical states in helium droplets.
A possible experiment to detect their presence has been
described by Close et al. \cite{close}. They
have suggested that alkali atoms, that
normally reside in a `dimple' on the surface
of $^4$He clusters \cite{scoles,anci1,ernst1},
may be drawn, when a vortex is present, {\it inside} the cluster 
along the vortex core. 
Spectroscopic experiments on the dopant atoms
could thus provide evidence of their existence,
since the line broadenings and shifts  would be different in the two
cases. We show in the following that alkali atoms are actually not suited
for such an experiment, but rather alkaline earth (Ca) atoms
may serve as probes to detect vortices.

Density Functional (DF) methods
\cite{evans} have become increasingly popular in recent years as
a useful computational tool to study the properties
of classical and quantum inhomogeneous fluids, especially for
large systems, for which they provide a good compromise
between accuracy and computational cost.
In particular, a quite accurate description of the properties of
inhomogeneous liquid $^4$He at zero temperature has been
obtained within a DF approach by using the energy
functional proposed in Ref. \cite{dupont} and later improved in
Ref. \cite{prica}. This later DF, which
has been successfully used over recent years
to study a variety of $^4$He systems like
clusters and films, is the one we use in the present work.

The minimization of the energy functional with
respect to density variations,
subject to the constraint of a given number of $^4$He atoms $N$,
leads to the equilibrium particle density profile $\rho ({\bf r})$,
thus allowing to study the  static properties
of the $^4$He system. When dynamical properties are studied (as
described in the following), we use the
Time-Dependent DF (TDDF) method
developed in Ref. \cite{giacomazzi}, which allows to obtain
both the $^4$He particle density $\rho ({\bf r},t)$ and
the velocity field ${\bf v}({\bf r},t)$.
Briefly, in the static (dynamic) case, one has
to solve a stationary (time-dependent)
non-linear Schr\"odinger-like equation for an
`order parameter' $\Psi ({\bf r})$ $(\Psi ({\bf r},t))$,
where the Hamiltonian
operator is given by $H=-\frac{\hbar }{2m}\nabla ^2+U[\rho,{\bf v}]$.
The effective potential $U$
is defined as the variational derivative of the
energy functional,
and its explicit expression is given in Ref. \cite{giacomazzi}. 
From the knowledge of
$\Psi\equiv \phi e^{i\Theta }$
one can get the
density $\rho({\bf r},t)=\phi^2$ and the fluid velocity field
${\bf v}({\bf r},t)=\frac{\hbar}{m}\nabla \Theta$.
To model the interaction of liquid $^4$He with foreign impurities we
use suitable He-impurity pair interaction potentials,
which will be described later on.

We work in 3D cartesian coordinates, and adopt the following procedure to
generate a vortex in the cluster in the most unbiased way.
We consider a cluster in a rotating frame of reference
with constant angular velocity $\omega_z$ around the z-axis \cite{don91}.
The Hamiltonian density $H$ then acquires an additional term
$-\omega_z \hat{L}_z$, $\hat{L}_z$ being the angular momentum
component along the z-axis.
We minimize $\Psi$ for this constrained Hamiltonian,
imposing $\Psi$ to be orthogonal, during
the minimization, to $\Psi _0=\sqrt{\rho _{eq} ({\bf r})}$
describing the minimum energy state of the vortex-free cluster;
we have applied the method to a $N=300$ droplet.
To generate a vortex line,
$\omega_z$ must be larger than a critical value -unknown in advance-
$\Omega_c = \Delta E/(N\hbar) $ \cite{dal96}, where $\Delta E$ is the energy
cost to create a vortex (which in the present case is about 70 K, see
Table \ref{table1}, and hence $\Omega_c \sim 3 \times 10^{10}$ s$^{-1}$),
 but not
so large that one could  generate a vortex  array \cite{don91}.
The particle density corresponding to this vortical configuration is
shown in Fig. 1(a). We have calculated the circulation of the velocity
field along a path enclosing the vortex core,
and have exactly found the value $N h /m$ appropriated for a quantized
vortex with $\langle \hat{L}_z \rangle=N\hbar $.

Note that since the vortex is quantized,
the vortical state is an eigenstate of the
angular momentum along the rotation axis, $\hat{L}_z$.
This means that our density profile
is the same as that one would obtain by using the
Feynman-Onsager {\it ansatz}, i.e. by adding to the energy functional
an extra centrifugal term associated with an order parameter
of the form $\sqrt{\rho }e^{i\Phi }$ ($\Phi$ being the azimuthal
angle), and finding the density profile by solving an equation
in the real quantity $\sqrt{\rho ({\bf r})}$.
This is the procedure used in the DF calculations of
Ref. \cite{barranco1}
to generate quantized vortex structures in helium drops
-and also in Bose-Einstein condensates of trapped gases \cite{dal96}-.
Instead, we have
not assumed a priori a quantized value for the total angular
momentum, but rather we generate
a fully quantized vortex state starting from a pure cluster.

To use atomic impurities as probes
of the presence of vortices in $^4$He drops,
ideally one would like to have an atom that is {\it barely} stable
on the surface of a pure drop, and becomes solvated
in its interior in the presence of a vortex.
The question of solvation vs. surface location for an
impurity atom in liquid $^4$He can be addressed
in an approximate way
within the model of Ref. \cite{lerner} where,
based on calculations of the energetics of impurities
interacting with liquid $^{4}$He, a simple criterion is proposed
to decide whether surface or solvated states are favored.
An adimensional parameter is defined
in terms of the impurity-He potential well depth
$\epsilon$ and the minimum position $r_m$,
$\lambda \equiv \rho \,\epsilon\, r_m/(2^{1/6}\sigma)$,
where $\rho$ and $\sigma$ are the bulk liquid density
and surface tension of $^4$He, respectively.
The criterion for solvation reads $\lambda > 1.9$
for the existence of solvated states \cite{lerner}.
One thus needs an impurity with $\lambda \sim 2$,
and such that its most stable state is on the drop surface.

Alkali atoms are known to have their stable state on the
surface of liquid $^4$He \cite{scoles,anci1}, and
they lie in the low $\lambda$ regime
($\lambda \sim 0.6-0.9$) \cite{lerner}. Accordingly,
for alkalis a surface state should always be preferred, even in the
presence of a vortex line.
We have verified this point considering Na and Rb, as
representative of light and heavy alkali, respectively.
The alkali-He interaction is of the
form proposed by Patil \cite{patil}.
We have compared the stable `dimple'
states of Na and Rb atoms on the surface of the $N=300$ cluster
hosting a vortex line,
with those of the same impurity trapped in the vortex core,
exactly at the cluster center. We have found that the latter
are energetically unfavored with respect to
surface states, see Table \ref{table1}.
It is worth to note that in the case of Na,
our results compare well with
the Path Integral Monte Carlo calculations of
Ref. \cite{nakayama}, where a binding energy of about
$\sim 7$ K is found for this cluster.
We also note that, unlike the case of strongly
bound impurities to $^4$He clusters \cite{barranco1},
which have their stable state inside
the cluster, and for which there exists a critical cluster size
below which the droplet+dopant+vortex complex is stable,
the alkalis cannot stabilize the vortex, whatever the droplet 
size is \cite{barranco1}.

There are other dopants, however, for which there is
clear evidence of a surface state on liquid $^4$He, 
i.e. alkaline earth atoms.
Absorption spectra of alkaline earth atoms (Ca, Ba and Sr)
attached to $^4$He clusters 
clearly support an outside location of Ca and Sr \cite{frank1},
and probably also of Ba \cite{frank2}.
To describe the He-impurity interaction
we employ an accurate ab-initio He-Ca pair potential
\cite{meyer}
used to study $^4$He$_N$+Ca droplets up to $N=75$ by
Diffusion Monte Carlo techniques \cite{marius}.
For such a potential, $\lambda \sim 2.2$, which apparently
indicates a solvated stable state. However,
for those cases where $\lambda $ is
close to the solvation threshold, consideration
of the shape of the potential
energy surface, as well as the well depth and equilibrium
internuclear distance, seems warranted \cite{scoles1}.

The stable state of a Ca atom in a $^4$He$_{300}$ vortex-free cluster
is shown in Fig. 1(b). 
Note that, in qualitative
agreement with the experimental evidence,
the `dimple'
appears to be much more pronounced than in the
case of alkalis \cite{giacomazzi}, reflecting the stronger He-atom interaction.
However, in the presence of a vortex, the stable state
is in the center of the cluster, as depicted in Fig. 1(c).
The surface state in this case is unstable
and, as the minimization proceeds, a Ca atom initially placed
on the surface near the vortex core, is gradually
drawn towards it and then sucked inside, eventually
reaching the stable state in the center of the drop.
The value of the angular momentum
for the converged $^4$He configuration is again  $\langle \hat{L}_z \rangle=N\hbar $.

The response of the impurity atom to
the different $^4$He environments shown in Fig. 1 might
be determined with spectroscopic measurements,
allowing to detect the presence of vortices:
the observed linewidths and shifts
of the excitation/emission spectra in the
two cases shown in Fig. 1 should be very different, reflecting the
`bubble' environment in one case [Fig. 1(c)], and
a more open environment in the other case [Fig. 1(b)].
Moreover, in the case of Fig. 1(b), bound-unbound transitions
should be observed with a significant probability,
thus implying a strong asymmetry in the observed spectra.

In the above picture one is assuming that  the vortex is long-lived
enough to allow a Ca atom, picked up randomly by the cluster, to
diffuse close to the top of the vortex core
and then to be drawn inside. We have no direct proof of the
stability of the cluster+vortex complex
on experimental time scales. However, we have indications that
the cluster+vortex+dopant complex should 
be stable at least on the
nanosecond time-scale. This conclusion comes from 
very long computing time simulations,
using the TDDF method \cite{giacomazzi},
to study the dynamics of the cluster+vortex+impurity complex.
During these simulations, the impurity atoms
were allowed to oscillate inside the `bubble'
in the cluster center [see  Fig.1(c)], and
the vortex line was always found to be stable,
without showing any tendency to
shrink, bend or migrate towards the surface of the
cluster.  

The solvated state for Ca in a
vortex-free cluster is a stationary but unstable configuration
against any displacement of the atom off the cluster center,
only a few K in energy above the stable, `dimple' state
(see Table \ref{table1}). 
This is a consequence of the borderline value of $\lambda $ for this
impurity, and implies that in a real experiment, a fraction
of Ca atoms might be trapped inside the clusters
for fairly long times, even in the absence of vortices.
For these atoms, the spectroscopic signals would be
similar to those coming from Ca atoms
trapped in the vortex core, making it
difficult to discriminate between the two cases.
A line shape calculation \cite{scoles} using as an input
the $^4$He density profiles around the impurity might help to
distinguish between solvated states of Ca with and without
vortex. Since the
extension of the method of Ref. \cite{scoles} -which is applied there to
the simpler case of the monoelectronic alkali atoms- is rather involved
for two-electron systems, we have not carried out such a
calculation. We instead suggest additional measurements which
may help to discriminate between the states shown in Fig. 1.

It appears from our calculations that the
energy of the impurity-cluster system is rather insensitive to the
location of Ca along the vortex core, once the atom is embedded in it.
Consequently,
vibrational modes of the impurity
along the vortex line are
expected to be soft. We have  confirmed this by TDDF calculations,
applying to the Ca atom a small
initial momentum in a given direction -radially, towards the
surface of the cluster for the `dimple' state of Fig. 1(b), and
along or perpendicular to the vortex line in
the case shown in Fig. 1(c)-.
We then let the impurity evolve in time, allowing for the
$^4$He environment to dynamically follow the atom motion while the
impurity oscillates around its equilibrium position.
This is done in practice by numerically solving 
by means of a discrete Verlet algorithm, 
as usually done in Molecular Dynamics calculations,
Newton's equation of motion
for the Ca atom under the
force due to the surrounding $^4$He liquid 

$$M \frac{d^2{\bf R}}{dt^2}=
-{\bf \nabla}_{{\bf R}} \left\{\int \rho({\bf r},t)V_{He-Ca}({\bf
R}(t)-{\bf r})d{\bf r}\right\}$$

In this expression $V_{He-Ca}$ is the pair potential describing the
He-impurity 
interaction, and the density $\rho({\bf r},t)$ is updated at each time step
according to the TDDF scheme for $^4$He \cite{giacomazzi}.  From
the  positions of the
Ca atom as a function of time,
relative to the center-of-mass of the Ca-He droplet system,
different frequencies characterizing
the impurity dynamics
can be found from a Fourier analysis of the calculated time series.

We report in Fig. 2 the calculated vibrational spectra.
The intensities are in arbitrary units, and normalized so that
the higher peak in each spectrum has unit height \cite{note1}.
It appears that the oscillation of Ca along the vortex core 
is indeed characterized by a single low frequency mode,
as compared with the more fragmented spectrum for vibrations perpendicular to
the vortex core. 
The presence of the soft mode is
a signature of solvation of a Ca atom inside the vortex. Indeed,
such a mode should be severely damped, or even absent, for
a solvated Ca atom in a vortex-free cluster, since
this configuration is unstable.
We also show for comparison  the 
vibration spectrum of a Ca atom in the `dimple' state
on the surface of a cluster without vortex. 
The peak just below 1 K is due to the  `dipolar'
vibration of the impurity inside the semi-spherical(spherical)
cavity in which it is trapped in the `dimple'(`bubble') state.
Additional peaks appear in the `dimple' spectra because of
the coupling of the Ca motion 
with the surface modes of the $^4$He nanodroplet,
which have similar frequencies
(for instance, the lowest energy, $l=2$ quadrupolar mode of a pure 
$^4$He$_{300}$
droplet occurs at $\sim 0.6$ K \cite{giacomazzi}).
All these modes lie in the microwave frequency regime and there are
experimental ideas to measure the 
corresponding vibrational frequencies \cite{ernst}. 

It is worth to see how  these modes are coupled, which is
particularly apparent when we displace the impurity
perpendicular to the vortex core (dashed line in Fig. 2).
To trigger this oscillation, we have given to the
the impurity a kinetic energy of about 2 K, three times as much
as in the other two cases. One may see that the spectrum displays
one peak corresponding to the `dipolar' mode discussed previously, and also
softer modes of characteristics similar to those found in the other
two cases. This coupling is possible because
the time evolution is adiabatic.  

Finally, we would like to emphasize that the dynamical behavior of
He-impurity systems depends in a sensitive way on the
details of the He-atom pair interaction. This calls for
improving the available 
interaction potentials to strengthen the scenario
described here, or to help in finding other atomic/molecular
impurities which may serve as probes of the presence
of vortices in $^4$He droplets.

We thank M.W.~Cole, F.~Dalfovo, W.~Ernst, S. Hern\'andez, G.~Scoles,
F.~Stienkemeier, and F.~Toigo for useful comments and discussions.
We thank Prof. W.~Meyer for the permission of using his unpublished
Ca-He potential.
F.A. acknowledges funding from MIUR-COFIN 2001 (Italy), and M.B. and M.P.
from BFM2002-01868 and 2001SGR00064 (Spain).

\bigskip

\centerline{Figure Captions}

Figure 1: From left to right, for a $^4$He$_{300}$ cluster,
atomic equidensity lines corresponding to:
(a) the vortical state;
(b) the deep `dimple' stable state of Ca, and
(c) the solvated stable state of Ca in the presence of
a quantized vortex.
Each box side has 45 ${\rm \AA}$ lenght.

Figure 2: Calculated frequency spectra for the
oscillations of a Ca atom in a $^4$He$_{300}$ cluster.
Solid line: solvated state [see Fig.1(c)], Ca
moving along the vortex core;
dashed line: solvated state, Ca moving perpendicular
to the vortex core;
dotted line: surface state [see Fig.1(b)], Ca moving in the radial
direction.

\pagebreak

\begin{table}

\caption{
Energies of several $^4$He$_{300}$ configurations.
In the left part of the Table, they are
referred to the total energy of the pure, vortex-free cluster
(-1384.5 K),
whereas in the right part they are
referred to that of the cluster+vortex configuration (-1313.4 K).
The configurations marked with an asterisk are
unstable stationary configurations.}
\begin{ruledtabular}
\begin{tabular}{lrclrc}

Configuration & E (K) & \vline    & Configuration & E (K) \\ \hline
Rb (center)$^*$ & +94.7 & \vline & Rb/vortex (center)$^*$ & +71.7 \\
Rb (top)        & -9.1  & \vline & Rb/vortex (top)        & -12.5 \\
Na (center)$^*$ & +64.3 & \vline & Na/vortex (center)$^*$ & +45.1 \\
Na (top)        & -8.1  & \vline & Na/vortex (top)        & -11.3 \\
Ca (center)$^*$ & -33.7 & \vline & Ca/vortex (center)     & -49.9 \\
Ca (top)        & -37.4 & \vline & Ca/vortex (top)   &  unstable  \\
Vortex          & +71.1 & \vline &                        &       \\
\end{tabular}
\end{ruledtabular}
\label{table1}
\end{table}


\begin{thebibliography}{99}
\bibitem{note1} 
We have added to the bare spectra a Lorentzian broadening
of width $\Delta \omega \sim 2\pi /t_{max}$, 
which reflects the finite duration of our simulations,
$t_{max}\sim 0.5 \,\times 10^{-9}$ s.
\end{thebibliography}
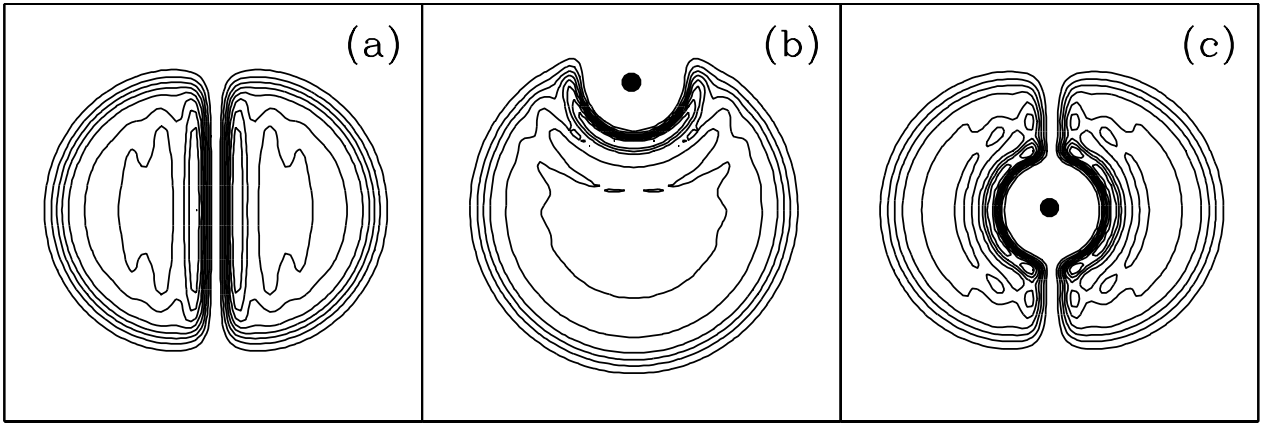
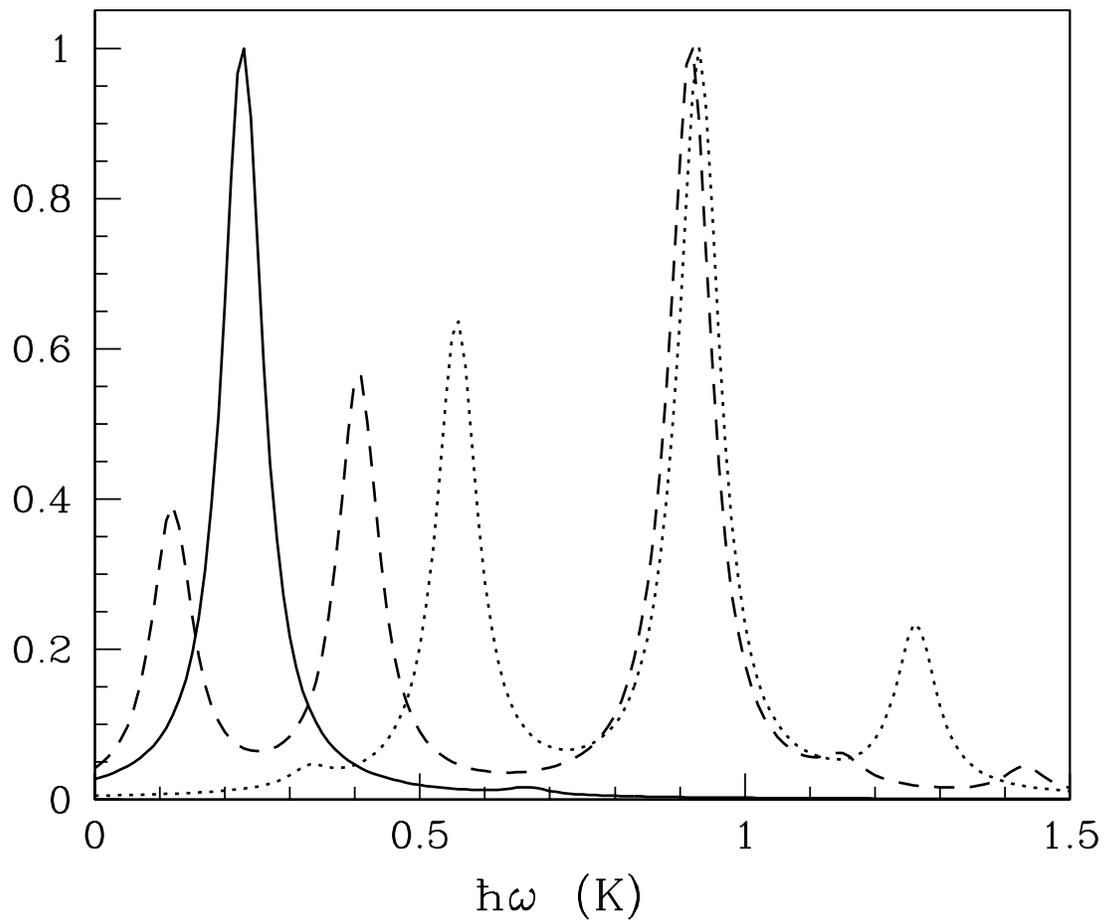
\end{document}